\documentstyle[amssymb,aps,12pt]{revtex}

\begin{document}
\draft
\author{O. B. Zaslavskii}
\address{Department of Mechanics and Mathematics, Kharkov V.N. Karazin's National\\
University, Svoboda\\
Sq.4, Kharkov 61077, Ukraine\\
E-mail: ozaslav@kharkov.ua}
\title{Exactly solvable models of two-dimensional dilaton cosmology with quantum
backreaction}
\maketitle

\begin{abstract}
We consider general approach to exactly solvable 2D dilaton cosmology with
one-loop backreaction from conformal fields taken into account. It includes
as particular cases previous models discussed in literature. We list
different types of solutions and investigate their properties for simple
models, typical for string theory. We find a rather rich class of everywhere
regular solutions which exist practically in every type of analyzed
solutions. They exhibit different kinds of asymptotic behavior in past and
future, including inflation, superinflation, deflation, power expansion or
contraction. In particular, for some models the dS spacetime with a
time-dependent dilaton field is the exact solution of field equations. For
some kinds of solutions the weak energy condition is violated independent of
a specific model. We find also the solutions with a singularity which is
situated in an infinite past (or future), so at any finite moment of a
comoving time the universe is singularity-free. It is pointed out that for
some models the spacetime may be everywhere regular even in spite of
infinitely large quantum backreaction in an infinite past.
\end{abstract}



\section{introduction}

Two-dimensional (2D) models of gravity attract much attention over last
decade. They capture the main non-trivial features inherent to 4D gravity
(existence of black holes and their evaporation, appearance of cosmological
singularities, etc.) and, due to their relative simplicity, enable to track
details which are obscured by mathematical complexity in 4D case. Stimulated
by investigations, started in 2D black hole physics \cite{callan}, at
present 2D dilaton gravity grew in a rather vast branch of semiclassical
gravity (see, e.g., the reviews\cite{dv}, \cite{od}). On the other hand,
although the 2D is much simpler than 4D one, a generic 2D model with
backreaction taken into account is not exactly solvable. Exactly solvable
models of 2D semiclassical dilaton gravity represent only special subset of
all possible 2D dilaton gravity theories and were studied mainly in the
context of black hole physics \cite{bil} - \cite{class}.

In the present paper we will be dealing with 2D dilaton cosmology accounting
for backreaction of quantum conformal fields. A particular solution for such
a system was found in \cite{maz} but it turned out that it suffers from the
presence of unavoidable physical singularities. The interest to exact
solutions in 2D semiclassical cosmology was further stimulated by the fact
that inclusion of backreaction in some another models does cure the problem
of cosmological singularities. This was observed in Ref. \cite{rey}, where
it was shown that this effect leads to smooth transition between
superinflation and FRW phases for some particular model and, thus, resolves
the problem of the graceful exit (brief review on this issue in the context
of dilaton theory can be found in \cite{rey2}). However, this result was
achieved at the cost of putting the quantum coupling parameter $\kappa =%
\frac{\hbar (N-24)}{24}$ to negative values. Meanwhile, the quasiclassical
approximation, within which all consideration was performed, implies $\hbar
\rightarrow 0$, $N\rightarrow \infty $, that is $\kappa >0$. The improved
model, free of unphysical restriction on $\kappa $, was analyzed in \cite{b1}%
, \cite{bcos} and was shown also to contain singularity-free solutions.

These studies on 2D dilaton cosmology with everywhere regular solutions
concerned only some special fixed models with the zero cosmological
constant. In the present paper we relax the demand of having the zero
cosmological constant (that, in the context of modern cosmology, looks more
physical) and (i) give the full list of cosmological solutions within the
exactly solvable models specified by the one relationship between the
coefficient of the gravitation-dilaton action, (ii) find among them examples
with everywhere bounded curvature.

In 4D world one can safely distinct between black hole and cosmological
horizons (for instance, in the de Sitter spacetime an observer, situating in
the static region, can observe that the area of a sphere is increasing,
while he is approaching the horizon). In 1+1 spacetimes, where there is only
one space dimension, such a distinction becomes conditional. To avoid
confusion between proper cosmological solutions and non-static regions
inside black holes, hereafter (unless otherwise stated explicitly) we will
be interested in such spacetimes, which are regular for all interval of
comoving time $-\infty <\tau <\infty $ (that is the most physically
interesting situation) or hit a singularity. If it happens at some finite $%
\tau =\tau _{0}$, the geometry is regular for $\tau _{0}<\tau <\infty $ or $%
-\infty <\tau <\tau _{0}$. Apart from this, we find also the intermediate
type of the metric when the singularity exists only in an infinite past (or
infinite future) with respect to a comoving observer. The solutions
discussed in \cite{kim} do not meet the above mentioned criteria and
represent rather a non-extreme black (or white) hole in its non-static
region, than a true cosmological solution.

The paper is organized as follows. In Sec. II we repeat briefly the main
formulas concerning exactly solvable models, following the approach of \cite
{class} (where a reader can find details), with modifications, necessary to
take into account that now we are dealing with homogeneous solutions instead
of static ones in \cite{class}. In Sec. III we list general solutions within
families of exactly solvable solutions and enumerate possible subclasses,
which are determined by the relationship between their parameters. In Sec.
IV we analyze in detail possible concrete types of solutions, which can be
obtained from generic ones by setting some parameters to zero, find explicit
exact solution and analyze their asymptotic behavior in far past and future.
Special emphasis is made on everywhere regular solutions, when the curvature
scalar remains bounded. In Sec. V we discuss the properties of generic types
of solutions and show, how results, obtained in Sec. IV for particular
cases, can be exploited to describe asymptotic behavior of generic exact
solutions. In Sec. VI, using some exact solutions, found in previous
sections, we show that even infinitely strong quantum backreaction may be
compatible with the regularity of cosmological solutions. In Sec. VII give
the summary of the results obtained.

\section{basic equations}

Consider the action 
\begin{equation}
I=I_{0}+I_{PL}\text{,}  \label{action}
\end{equation}
where 
\begin{equation}
I_{0}=\frac{1}{2\pi }\int_{M}d^{2}x\sqrt{-g}[F(\phi )R+V(\phi )(\nabla \phi
)^{2}+U(\phi )]\text{,}  \label{clac}
\end{equation}
the Polyakov-Liouville action \cite{pl} $I_{PL}$, incorporating effects of
Hawking radiation and its back reaction on the black hole metric, can be
written as 
\begin{equation}
I_{PL}=-\frac{\kappa }{2\pi }\int_{M}d^{2}x\sqrt{-g}[\frac{(\nabla \psi )^{2}%
}{2}+\psi R]\text{.}  \label{PL}
\end{equation}
The function $\psi $ obeys the equation 
\begin{equation}
\Box \psi =R\text{,}  \label{psai}
\end{equation}
where $\Box =\nabla _{\mu }\nabla ^{\mu }$, $\kappa =N/24$ is the quantum
coupling parameter, $N$ is number of scalar massless fields, $R$ is a
Riemann curvature. We omit the boundary terms in the action that do not
affect the form of field equations.

Varying the action with respect to a metric gives us $(T_{\mu \nu }=2\frac{%
\delta I}{\delta g^{\mu \nu }})$ 
\begin{equation}
T_{\mu \nu }\equiv T_{\mu \nu }^{(0)}-T_{\mu \nu }^{(PL)}=0\text{,}
\label{6}
\end{equation}
where 
\begin{equation}
T_{\mu \nu }^{(0)}=\frac{1}{2\pi }\{2(g_{\mu \nu }\Box F-\nabla _{\mu
}\nabla _{\nu }F)-Ug_{\mu \nu }+2V\nabla _{\mu }\phi \nabla _{\nu }\phi
-g_{\mu \nu }V(\nabla \phi )^{2}\}\text{,}  \label{7}
\end{equation}
\begin{equation}
T_{\mu \nu }^{(PL)}=\frac{\kappa }{2\pi }\{\partial _{\mu }\psi \partial
_{\nu }\psi -2\nabla _{\mu }\nabla _{\nu }\psi +g_{\mu \nu }[2R-\frac{1}{2}%
(\nabla \psi )^{2}]\}\text{.}  \label{8}
\end{equation}

Variation of the action with respect to $\phi $ gives rise to the equation 
\begin{equation}
R\frac{dF}{d\phi }+\frac{dU}{d\phi }=2V\Box \phi +\frac{dV}{d\phi }(\nabla
\phi )^{2}  \label{9}
\end{equation}

In general, field equations cannot be solved exactly and the function $\psi $%
, the dilaton $\phi $ and metric depend on both time-like ($t$) and
space-like ($x$) coordinates: $\psi =\psi (t$,$x)$, $\phi =\phi (t$, $x)$.
In what follows we restrict ourselves to such kind of solutions that $\psi $
can be expressed in terms of $\phi $ only: $\psi =\psi (\phi )$. This leads
to the existence of the Killing vector which, in our case is assumed to be a
spacelike instead of timelike in \cite{class}.

In general, eqs. (\ref{6}) - (\ref{9}) cannot be solved exactly even under
these restrictions. However, there exist a family of models which is exactly
solvable. For these models the action coefficients are not arbitrary but
satisfy the relationship 
\begin{equation}
V=\omega (u-\frac{\kappa \omega }{2})+C(u-\kappa \omega )^{2}\text{,}
\label{exact}
\end{equation}
where $C$ is a constant, $u\equiv \frac{dF}{d\phi }$, $\omega \equiv \frac{%
d\ln \left| U\right| }{d\phi }$ \cite{exact}. The full list of possible
types of solutions within the family (\ref{exact}) was suggested in \cite
{class}. Thus, instead of looking for cosmological solutions all over again,
we may simply borrow them from \cite{class}. Being trivial from the formal
viewpoint (one only needs to interchange temporal and spatial coordinates
and signs in some folrmulas), this procedure gives rise, however, to
qualitatively another types of solutions in the sense that every static
solution has its homogeneous counterpart, and vice versa. Not all of them
are of physical interest, but, as we will see below, corresponding exactly
solvable models contain cases which can be interpreted as describing
evolution of 2D universe.

It is convenient to work in the conformal gauge 
\begin{equation}
ds^{2}=g(-dt^{2}+d\sigma ^{2})  \label{g}
\end{equation}
where, in accordance with the choice of the Killing vector, $g=g(t)$ and
does not depend on a space-like coordinate $\sigma $. In the gauge (\ref{g})
the curvature 
\begin{equation}
R=g^{-1}(\dot{g}/g\dot{)}=2a^{-1}\frac{\partial ^{2}a}{\partial \tau ^{2}}%
\text{, }g\equiv a^{2}\text{,}  \label{R}
\end{equation}
$a$ is a scale factor, $\tau $ is a comoving time. (Throughout the paper dot
denotes differentiation with respect to a time-like coordinate $t$, prime -
with respect dilaton $\phi .$) For exactly solvable models \cite{class} 
\begin{equation}
\psi =\psi _{0}+\gamma t\text{,}  \label{psg}
\end{equation}
where 
\begin{equation}
\psi _{0}=\eta +2CH\text{, }\eta \equiv \int d\phi \omega \text{.}
\label{psi}
\end{equation}
It follows from (\ref{psai}) that 
\begin{equation}
g=e^{-\psi -bt}=e^{-\psi _{0}-\delta t}\text{,}  \label{gps}
\end{equation}
where $b$ is a constant, $\delta =\gamma +b$.

It is instructive to write down also formulas in the isotropic coordinates $%
x^{\pm }=t\pm x$. Then 
\begin{equation}
ds^{2}=-e^{2\rho }dx^{+}dx^{-}\text{.}  \label{iso}
\end{equation}
The Polyakov-Liouville stress-energy tensor 
\begin{eqnarray}
T_{\pm \pm }^{(PL)} &=&\frac{2\kappa }{\pi }[\partial _{\pm }^{2}\rho
-(\partial _{\pm }\rho )^{2}-t_{\pm }(x^{\pm })]\text{,}  \label{stress} \\
T_{+-}^{(PL)} &=&-\frac{2\kappa }{\pi }\partial _{+}\partial _{-}\rho \text{,%
}  \nonumber
\end{eqnarray}
the curvature 
\begin{equation}
R=8e^{-2\rho }\partial _{+}\partial _{-}\rho \text{.}
\end{equation}
Provided the metric $g$ and $\psi $ depend on $t$ only, it follows from (\ref
{8}) and (\ref{psg}) that the quantity $T_{01}^{(PL)}=0$. Then (\ref{stress}%
) it entails that for spacetimes under discussion $t_{+}=t_{-}=const$.
Comparing (\ref{8}), (\ref{gps}) and (\ref{iso}), we see that 
\begin{equation}
t_{+}=-\frac{b^{2}}{16}\equiv -\frac{(\delta -\gamma )^{2}}{16}\text{,}
\end{equation}
where we put $b=\delta -\gamma $ to retain succession with notations of \cite
{class}.

\section{general case, types of solutions}

It turns out \cite{class} that the solutions of field equations can be
represented in the form: 
\begin{equation}
CH(\phi )=-\frac{\delta t}{2}+\ln \left| f\right| \text{,}  \label{hf}
\end{equation}
\begin{equation}
H=F-\kappa \int d\phi \omega =F-\kappa \ln \left| U\right| \text{,}
\label{h}
\end{equation}
where the function $f$ obeys the equation

\begin{equation}
\frac{d^{2}f}{dt^{2}}=f\varepsilon ^{2}\text{, }\varepsilon ^{2}=\frac{%
\delta ^{2}}{4}+\alpha C\text{,}
\end{equation}
\begin{equation}
\alpha =\frac{\kappa \gamma (2\delta -\gamma )}{2(1-2\kappa C)}\text{.}
\label{alpha}
\end{equation}
If $U\equiv 0$, the second formula in (\ref{h}) loses its meaning but other
ones retain their validity, where the function $\omega $ should be
understood as an arbitrary function, parametrizing a solution.

We get the following different cases.

I$_{a}$. $\varepsilon ^{2}>0$, $f=\frac{sh\varepsilon t}{\varepsilon }$; I$%
_{b}$: $f=\frac{ch\varepsilon t}{\varepsilon }$;

II$_{a}$: $\varepsilon =0$, $f=t$; II$_{b}$: $f=1$;

III: $\varepsilon ^{2}\equiv -\varkappa ^{2}<0$, $f=\frac{\sin \varkappa t}{%
\varkappa }$.

Let us write the potential as 
\begin{equation}
U=\Lambda \exp (\int d\phi \omega )\text{.}  \label{u}
\end{equation}
Then one can derive from field equations that 
\begin{equation}
\frac{\Lambda C}{1-2\kappa C}=z\text{,}
\end{equation}
where $z=-1$ for the I$_{b}$ case, $z=0=\Lambda $ for II$_{b}$ and $z=1$ in
cases I$_{a}$, II$_{a}$, III.

The Riemann curvature reads the following.

I$_{a}$, II$_{a}$, III: 
\begin{equation}
R=\frac{UC}{1-2\kappa C}[2+\frac{\omega }{CH^{\prime }}-\frac{1}{%
C^{2}H^{\prime }}\left( \frac{\omega }{H^{\prime }}\right) ^{\prime }q^{2}]%
\text{.}  \label{gcur}
\end{equation}
I$_{b}:$ 
\begin{equation}
R=\frac{UC}{1-2\kappa C}[2+\frac{\omega }{CH^{\prime }}+\frac{1}{%
C^{2}H^{\prime }}\left( \frac{\omega }{H^{\prime }}\right) ^{\prime }q^{2}]
\label{r2}
\end{equation}
II$_{b}$: 
\begin{equation}
R=-\frac{e^{\eta }}{1-2\kappa C}\frac{1}{C^{2}H^{\prime }}\left( \frac{%
\omega }{H^{\prime }}\right) ^{\prime }\frac{\delta ^{2}}{4}\text{.}
\label{r3}
\end{equation}

Here $q=(\frac{df}{dt}-\frac{\delta }{2}f)$.

In a similar way, we get the general structure of the expression for quantum
stresses. Two nonzero components of quantum stresses are connected for
conformal fields by the well known relationship $T_{0}^{0(PL)}+T_{1}^{1(PL)}=%
\frac{\kappa R}{\pi }$ (see eq. (\ref{8})). Here we list the component $%
T_{0}^{0(PL)}$only. One obtains from (\ref{8}), (\ref{psg}), (\ref{psi}), (%
\ref{alpha}): 
\begin{eqnarray}
T_{0}^{0(PL)} &=&\frac{1}{4\pi g}[\kappa (\frac{\partial \psi _{0}}{\partial
t}+2\delta )\frac{\partial \psi _{0}}{\partial t}+2\alpha (1-2\kappa C)]%
\text{,}  \label{t00a} \\
\frac{\partial \psi _{0}}{\partial t} &=&(\frac{\omega }{CH_{\phi
}^{^{\prime }}}+2)\frac{q}{f}\text{,}  \nonumber
\end{eqnarray}
whence 
\begin{equation}
T_{0}^{0(PL)}=\frac{\kappa }{4\pi }\frac{\left| UC\right| }{1-2\kappa C}Z%
\text{,}  \label{t00}
\end{equation}

\begin{equation}
Z=(\frac{q\omega }{CH^{^{\prime }}}+2\dot{f})^{2}-(\delta -\gamma )^{2}f^{2}%
\text{,}  \label{z}
\end{equation}
except the case II$_{b}$, when 
\begin{equation}
T_{0}^{0(PL)}=\frac{\kappa }{4\pi }e^{\eta }Z\text{,}
\end{equation}
$Z=\frac{\delta ^{2}}{4}\left( \frac{\omega }{CH^{^{\prime }}}\right)
^{2}-(\delta -\gamma )^{2}$.

\section{Particular cases and limiting transitions}

The solutions obtained depend on several parameters. In what follows it is
assumed that the dilaton is not identically constant. The quantities $%
\Lambda $ and $C$ enter the definition of the action coefficients: $\Lambda $
is the ''amplitude'' of the potential $U$ of a generic model according to
eq. (\ref{u}), while the parameter $C$ defines the coefficient $V$ of an
exactly solvable one (\ref{exact}). Meanwhile, the quantities $\delta $ and $%
\alpha $ are the parameters of the solutions of field equations, they do not
enter the action but characterize the different solutions for the same
model. Let us denote the symbolically [$C$, $\Lambda $]( $\delta $, $\alpha $%
) the solutions with given parameters for a given action, where it is
supposed that the values of parameters differ from zero, unless otherwise
stated explicitly. Consider first the case

\subsection{$C=0$}

Now 
\begin{equation}
\psi _{0}=\eta \text{, }g=e^{-\eta -\delta t}\text{.}  \label{c=0}
\end{equation}

In the cases [$0$, $0$]($0$, $\alpha $) and [$0$, $0$]($\delta $, $0$) it
turns out that field equations are mutually inconsistent, so these cases
cannot be realized.

\subsubsection{Type [$0$, $0]$($0$, $0$)}

\begin{eqnarray}
H &=&At\text{, }g=e^{-\eta }\text{, }R=-\frac{A^{2}}{H^{\prime }}\left( 
\frac{\omega }{H^{\prime }}\right) ^{\prime }e^{\eta }\text{,}  \label{000}
\\
\text{ }T_{0}^{0(PL)} &=&\frac{\kappa }{4\pi }A^{2}\frac{\omega ^{2}e^{\eta }%
}{H^{^{\prime }2}}>0\text{.}  \nonumber
\end{eqnarray}
Here $A$ is an arbitrary constant.

Below we will mainly concentrate on potentials of the form 
\begin{equation}
H=e^{-2\phi }  \label{d0}
\end{equation}
and 
\begin{equation}
H=e^{-2\phi }-\kappa d\phi \text{, }d>0\text{.}  \label{d}
\end{equation}
The condition $d>0$ ensures that $H^{\prime }(\phi )<0$ and does not change
the sign.

Example.

Let , for definiteness, $A<0$, 
\begin{equation}
\omega =-2n,0<n<2\text{.}  \label{ex}
\end{equation}
Then for the model (\ref{d0}) we have superinflaiton with the exact solution 
\begin{equation}
a\backsim (-\tau )^{p}\text{, }p=-\frac{n}{2-n}\text{, }R\backsim \frac{1}{%
\tau ^{2}}\text{, }\phi =\phi _{0}+\frac{1}{n-2}\ln \left| \tau \right| 
\text{,}  \label{p}
\end{equation}
($\phi _{0}$ is a constant). Hereafter we use for shortness dimensionless
time $\tau $. The case $n=2$ corresponds to de Sitter spacetime with $%
a\backsim \exp (\tau )$.

More interesting situation arises for the case (\ref{d}). Then for $\tau
\rightarrow -\infty $ eqs. (\ref{p}) hold asymptotically ($\phi \rightarrow
-\infty $), whereas for $\tau \rightarrow \infty $ ($\phi \rightarrow \infty 
$) we have 
\begin{equation}
a\backsim \tau \text{, }R\backsim \tau ^{-s},s=\frac{2n+2}{n}\text{.}
\label{ar}
\end{equation}
Thus, we have a graceful exit from superinflation to the FRW phase. This
solution generalizes previous result of \cite{bcos} (which holds for $n=1$)
to the case of an arbitrary $0<n<2.$

Let now $n=2$. Then for the model of the type (\ref{d}) $a\backsim \exp
(\tau )$, $\phi \backsim \frac{\tau }{2}\rightarrow -\infty $ at $\tau
\rightarrow -\infty $ and $a\backsim \tau $ and $R\backsim \tau ^{-3}$ at $%
\tau \rightarrow \infty $. Thus, we obtain the graceful exit from inflation
to FRW (Friedmann-Robertson-Walker) phase.

For an arbitrary $n$ the curvature 
\begin{equation}
R\backsim \frac{\exp [-(2n+2)\phi ]}{[2\exp (-2\phi )+\kappa d]^{3}}
\end{equation}
remains finite everywhere for any $1\leq n\leq 2$.

It is instructive to compare this with previous papers on semiclassical
exactly solvable cosmological models. Actually, the solution considered in 
\cite{kim} corresponds just only to the case $[0,0](0,0)$and particular
model (\ref{d}) for $n<0$. In \cite{kim} there was obtained the power
asymptotic behavior at the beginning of expansion near $\tau =0$, where $%
a\backsim \tau $ that it is rather difficult to call ''inflation'' (as it
was called in \cite{kim}). Moreover, as the spacetime is geodesically
incomplete in the case considered there (the parameter $\tau $ having
meaning of comoving time is finite), it seems that this is not a
cosmological solution at all but, rather, it describes the transition from a
homogeneous to a static, when an observer crosses an event horizon of a
white hole where $a=g=0$. Therefore, the finiteness of the curvature on the
horizon tells nothing about possible existence of the singularity which may
lie behind the horizon.

Compare now our results with those in \cite{bcos}. Using $a=\exp (\phi )$
from (\ref{000}), we obtain (taking for simplicity in (\ref{000}) $A=-1$)
for the model (\ref{d}) 
\begin{equation}
\tau =\kappa de^{\phi }-2e^{-\phi }=\kappa da-2a^{-1}\text{.}  \label{tf}
\end{equation}
Solving eq. (\ref{tf}) with respect to $e^{\phi }$, we obtain 
\begin{equation}
a=\frac{1}{\kappa d}(\sqrt{\tau ^{2}+2\kappa d}+\tau )\text{.}  \label{a1}
\end{equation}
This corresponds to eq. (17) of \cite{bcos} (a reader should bear in mind
that in \cite{bcos} by definition $\kappa =\frac{N\hbar }{12}$, while we use 
$\kappa =\frac{N\hbar }{24}$). Choosing the values of constants in \cite
{bcos}, $\alpha =1$, $\beta =0$ (this can be always done without loss of
generality by a proper shift and recalling of the variable $\tau $) and $d=1$
in our paper, we achieve full coincidence. The curvature is everywhere
bounded. For $\tau \rightarrow \infty $ $R\backsim \tau ^{-4}$ that agrees
with (\ref{ar}) in the case $n=1$.

It is also worth commenting on some general properties of the geometry and
quantum stresses for the kind of solution under discussion, generalizing the
previous observations made for particular representatives of the models
studied in \cite{bcos}, \cite{kim2}. If $\omega =const$ (that is indeed the
case for the most popular dilaton potentials and was used in the above
consideration), $R=\frac{A^{2}\omega H^{\prime \prime }e^{\eta }}{H^{\prime
3}}$. For the nonsingular models, which represent the most interest, $%
H^{\prime }$ does not vanish. If, for definiteness, $H^{\prime }<0$ and $%
\omega <0$, we obtain that $R>0.$ Taking into account (\ref{R}), we obtain
that in {\it any} kind of such a model $\frac{d^{2}a}{d\tau ^{2}}>0$: the
universe is ever accelerating.

The WEC (weak energy condition) is violated ($T_{0}^{0(PL)}>0$) for {\it any 
}kind of the solution $[0,0](0,0)$. On the other hand, it turns out that for
any solution of the type $[0,0](0,0)$ the classical part of the energy
density vanishes. Indeed, let us write down field equations in the form 
\begin{equation}
T_{cl.\mu }^{\nu }=T_{\mu }^{\nu (PL)}+T_{q\mu }^{\nu (\phi )}\equiv \theta
_{\mu }^{\nu }\text{,}
\end{equation}
where the term $T_{q\mu }^{\nu (\phi )}$ is obtained by varying the part of
the term in the gravitation-dilaton part that depends on $\kappa $
explicitly. By definition, the term $T_{cl.\mu }^{\nu }$, when expressed in
terms of the metric and dilaton according to (\ref{6}), does not contain $%
\kappa $. Then, calculating (\ref{6}) with the condition of exact
solvability (\ref{exact}) taken into account, we obtain that $\theta
_{0}^{0}=0$.

\subsubsection{Type $[0,0](\delta ,\alpha )$}

\begin{eqnarray}
&&H=\frac{\alpha }{\delta }t\text{, }a=\exp (-\frac{\eta }{2}-\frac{\delta
^{2}}{2\alpha }H)\text{, }R=-\frac{\alpha ^{2}}{\delta ^{2}}\frac{1}{%
H^{\prime }}\left( \frac{\omega }{H^{\prime }}\right) ^{\prime }\exp (\eta +%
\frac{\delta ^{2}}{\alpha }H)\text{,}  \label{0011} \\
&&T_{0}^{0(PL)}=\frac{\exp (\eta +\frac{\delta ^{2}H}{\alpha })}{4\pi }%
\alpha [\kappa (\frac{\alpha \omega }{\delta H^{^{\prime }}}+2\delta )\frac{%
\omega }{\delta H^{^{\prime }}}+2]\text{.}  \nonumber
\end{eqnarray}

If $\alpha =A\delta $ and $\delta \rightarrow 0$, while $A$ is kept fixed, (%
\ref{0011}) turns into (\ref{000}). If $H^{\prime }<0$ $H^{\prime \prime }>0$
and $\omega <0$, the curvature $R>0$, so similarly to the $[0,0](0,0)$ case
we obtain accelerating spacetimes only.

If $\delta >0$ and $\alpha >0$ (this is just the case discussed in \cite
{bcos}) the curvature diverges at $t\rightarrow \infty $. Let now $\delta >0$%
, $\alpha \equiv -\kappa \mu <0$. Then for the model (\ref{d}) with $\omega
=-2n$ the curvature $R$ is bounded everywhere if $q\equiv 2n+2-2p$ $\geq 0$,
where $p\equiv \frac{d\delta ^{2}}{2\mu }$, otherwise it diverges at $\phi
\rightarrow \infty $.

Example.

For the model (\ref{d}) we obtain at $t\rightarrow -\infty $: $a\backsim
(-\tau )\rightarrow \infty $, $\exp (-2\phi )\backsim (-t)\rightarrow \infty 
$, $R\backsim \left| \tau \right| ^{-2}\left( \ln \left| \tau \right|
\right) ^{-2}$. Thus, $R$ always remains bounded at $\tau \rightarrow
-\infty $ independent of the value of $q$.

At $t\rightarrow \infty $ $\phi \backsim \frac{\mu }{d\delta }t$, $a\backsim
\exp [t\frac{\mu }{\delta d}(1-p)]$. If $p\leq 1$, the spacetime is
geodesically complete since $\tau \rightarrow \infty $. Consider first the
case $p<1$. Then $a\backsim \tau $, , $\phi \backsim \ln \tau $, $R\backsim
\exp (-q\phi )\backsim \exp (-\frac{\mu }{\delta d}qt)\backsim \tau ^{-s}$, $%
s=q(1-p)^{-1}>0$. Thus, we have FRW-like expansion. Let now $p=1$. Then $%
a\rightarrow 1$ and the spacetime becomes asymptotically flat.

The above solutions $[0,0](0,0)$ and $[0,0](\delta ,\alpha )$ represent what
was called, correspondingly, ''first and second branches'' in \cite{bcos}.

For the model (\ref{d}) with $\omega =-2$ it is convenient to rewrite the
scale factor in (\ref{0011}) in the form 
\begin{equation}
a=\exp (\phi -\frac{\delta t}{2})=\exp [\phi (1+\frac{\delta ^{2}d\kappa }{%
2\alpha })-\frac{\delta ^{2}}{2\alpha }e^{-2\phi }]\text{.}  \label{a11}
\end{equation}
Introducing notation $\xi =e^{-2\phi }$ and choosing the values of constants 
$\delta ^{2}=2\mu >0$, $\alpha =\kappa \mu $, $d=1$ we have 
\begin{equation}
a=\xi ^{-1}\exp (-\frac{\xi }{\kappa })
\end{equation}
that coincides in this particular case with eq. (24) of \cite{bcos}.

It is worth stressing that the choice of the sign of $\alpha $ (while
keeping $\delta $ positive) has a crucial effect on the asymptotic behavior
of the geometry$.$ For $\alpha >0$ the solution $[0,0](\delta ,\alpha )$
contains the singularity in agreement with (\cite{bcos}) whereas for $\alpha
<0$, as we saw above, the geometry is everywhere regular.

\subsubsection{Type $[0,\Lambda ](\delta ,\alpha )$}

\begin{eqnarray}
H= &&h(t)=H_{0}+\frac{\alpha }{\delta }t+De^{-\delta t}\text{, }R=\frac{%
e^{\eta }}{H^{^{\prime }}}[\omega \Lambda -\left( \frac{\omega }{H_{\phi
}^{\prime }}\right) _{\phi }^{\prime }(\frac{\alpha ^{2}}{\delta ^{2}}%
e^{\delta t}-2\alpha D+D^{2}\delta ^{2}e^{-\delta t})]\text{, }  \label{0111}
\\
&&T_{0}^{0(PL)}=\frac{1}{4\pi }e^{\eta +\delta t}\{2\alpha +\kappa (\frac{%
\alpha }{\delta }-D\delta e^{-\delta t})\omega H^{^{\prime }-1}[2\delta +(%
\frac{\alpha }{\delta }-D\delta e^{-\delta t})\omega H^{^{\prime }-1}]\}%
\text{,}  \nonumber
\end{eqnarray}
\begin{equation}
D\delta ^{2}=-\Lambda .  \label{qdl}
\end{equation}

Let $\Lambda <0$, $D>0$, $\delta <0$, $\alpha >0$. By a proper shift in $t$,
one can always achieve $D=1$. Then there exists some point $t_{0}$ at which $%
h^{\prime }(t_{0})=0$. Let $H=e^{-2\phi }$, $\omega =-2$. Then at $%
t\rightarrow \infty $, $\phi \rightarrow -\infty $ we have a flat spacetime, 
$\phi =-\frac{\left| \delta \right| t}{2}$, $g\rightarrow 1$, $R\thicksim
\kappa \phi e^{2\phi }\rightarrow 0$. Consider $t\rightarrow t_{0}$, where $%
h^{\prime }(t_{0})=0$ and choose $H_{0}=\frac{\alpha }{\delta ^{2}}(\ln 
\frac{\alpha }{\delta ^{2}}-1)$, so that $h(t_{0})=0$, $h=\alpha \frac{%
(t-t_{0})^{2}}{2}+...$ near $t_{0}$. Then $\phi \rightarrow \infty $, $%
g\rightarrow \infty $, the proper time $\tau \thicksim \ln
(t-t_{0})\rightarrow -\infty $, $a\backsim \exp (-\tau )$, $\phi \backsim
-\tau $. We obtain de Sitter space undergoing deflation.

There also exists the second branch of solutions: $-\infty <t<t_{0}$. For
this branch we have inflation in future: $a\backsim \exp (\tau )\rightarrow
\infty $. In infinite past $\exp (-2\phi )\backsim \frac{\alpha }{\left|
\delta \right| }\left| t\right| $, $a\backsim -\tau $, $R\backsim \tau
^{-2}\ln ^{-2}\left| \tau \right| $. The point $\tau =0$ is singular.

If $\alpha >0$, $\delta >0$, the universe in the second branch starts from
the flat state and ends up with eternal inflation.

\subsubsection{Type $[0,\Lambda ](\delta ,0)$}

\begin{equation}
H=H_{0}+De^{-\delta t}\text{, }g=e^{-\eta -\delta t}\text{, }R=\frac{U}{%
H^{^{\prime }}}[\omega +(H-H_{0})\left( \frac{\omega }{H^{\prime }}\right)
^{\prime }]\text{,}  \label{hd}
\end{equation}
\begin{equation}
T_{0}^{0(PL)}=\frac{\kappa \omega U}{4\pi H^{^{\prime }}}[2-\frac{\omega
(H-H_{0})}{H^{^{\prime }}}]\text{.}
\end{equation}

Example. 
\begin{equation}
H=Ae^{-\phi }+e^{-2\phi }\text{, }A>0.  \label{a0}
\end{equation}
Let $\delta =-\left| \delta \right| <0$, $H_{0}\equiv -e^{\left| \delta
\right| t_{0}}<0$, $D>0$, $\Lambda <0$, $\omega =-2$. We may achieve $D=1$.
Then $t_{0}<t<\infty $. When $t\rightarrow t_{0}$ $\phi \rightarrow \infty $%
, $a\backsim (t-t_{0})^{-1}\backsim \exp (-\tau ),\tau \rightarrow -\infty $%
. At future $\tau \rightarrow \infty $, $\phi =-\frac{1}{2}\left| \delta
\right| t\rightarrow \infty $, $a\rightarrow 1$. Thus, we have graceful exit
from deflation to the flat spacetime.

In the particular case $A=2$ it is possible to obtain the explicit solution
for the whole region$.$ Solving (\ref{hd}) with respect to $e^{-\phi }$, we
find 
\begin{equation}
e^{-\phi }=\sqrt{1+H_{0}+De^{-\delta t}}-1\text{,}
\end{equation}
\begin{equation}
a=\frac{\exp (-\frac{\delta t}{2})}{\sqrt{1+H_{0}+De^{-\delta t}}-1}\text{.}
\end{equation}

The solution $[0$, $\Lambda ]$ $(0$, $0$) proves to be inconsistent with
field equations.

\subsubsection{Type [$0$, $\Lambda $]($0$, $\alpha $)}

\begin{eqnarray}
&&H=\frac{\alpha t^{2}}{2}+H_{0}\text{, }g=e^{-\eta }\text{, }\Lambda
=-\alpha =\frac{\kappa \gamma ^{2}}{2}>0\text{, }  \label{0101} \\
&&R=\frac{U}{H^{\prime }}[\omega +2\left( \frac{\omega }{H^{\prime }}\right)
^{\prime }(H-H_{0})]\text{,}  \nonumber \\
&&T_{0}^{0(PL)}=-\frac{U}{2\pi }[1+\left( \frac{\omega }{H^{\prime }}\right)
^{2}\kappa (H-H_{0})]\text{.}  \nonumber
\end{eqnarray}
This kind of solutions exists due to quantum effects ($\kappa \neq 0$) only.

Examples.

$H=e^{2\phi }-Ae^{-2\phi }$, $\omega =-2$, $A>0$. Then we have the solution
which behaves asymptotically as $a\backsim \exp (\tau )$ at $\tau
\rightarrow -\infty $ and $a\backsim \exp (-\tau )$ at $\tau \rightarrow
\infty $. Thus, we have the everywhere regular solution which starts at the
inflation phase, passes through the maximum value of $a$ and ends up with
deflation.

$H=e^{\phi }-ce^{-3\phi }$, $\omega =-2$, $c>0$. Then at $\tau \rightarrow
-\infty $ $a\backsim (-\tau )^{-2}$ and at $\tau \rightarrow \infty $ $%
a\backsim \tau ^{-2}$. Thus, superinflation is changed to superdeflation
with everywhere bounded curvature.

\subsection{$C\neq 0$}

\subsubsection{ Type $[C,\Lambda ](0,0)$.}

Then 
\begin{eqnarray}
H &=&H_{0}+C^{-1}\ln \left| \frac{t}{t_{0}}\right| \text{, }g=e^{-\eta }(%
\frac{t_{0}}{t})^{2}=\frac{const}{\left| U\right| t^{2}}\text{, }a=\frac{%
a_{0}}{\left| t\right| \sqrt{\left| U\right| }}\text{,}  \label{1100} \\
\text{ }R &=&\frac{U}{1-2\kappa C}[\frac{\omega }{H^{\prime }}+2C-\left( 
\frac{\omega }{H^{\prime }}\right) ^{\prime }\frac{1}{H^{\prime }C}]\text{,}
\\
\Lambda C &=&(1-2\kappa C)t_{0}^{-2}>0\text{,}  \nonumber
\end{eqnarray}
\begin{equation}
T_{0}^{0(PL)}=\frac{\kappa UC}{4\pi (1-2\kappa C)}(2+\frac{\omega }{%
CH^{^{\prime }}})^{2}>0\text{,}
\end{equation}
where $a_{0}$, $t_{0}$ are constants. Thus, the weak energy condition (WEC)
is always violated for this type of solutions.

It is worth noting that the dependence of the metric on the dilaton $g(\phi
) $ for this kind of solutions coincides with that for $[0,0](\delta ,\alpha
)$, provided $C=\delta ^{2}/2\alpha $.

On the other hand, the dependence of the dilaton on the proper time actually
coincides with that for the $[0,0](0,0)$ since in both cases 
\begin{equation}
\tau (\phi )=const\int d\phi H^{\prime }(\phi )\exp (-\eta /2).
\end{equation}

Consider the model (\ref{d0}) with $H_{0}=0$, $U=\Lambda e^{-4\phi }$, $%
\Lambda $, $C>0$, $-\infty <t<t_{0}$. Then 
\begin{equation}
\tau =-\tau _{0}\ln (\ln \left| \frac{t}{t_{0}}\right| )\text{, }
\end{equation}
\begin{equation}
a=a_{1}\exp [-\exp (-\frac{\tau }{\tau _{0}})+\frac{\tau }{\tau _{0}}]\text{%
, }a_{1}=\frac{a_{0}C}{\left| t_{0}\right| \sqrt{\Lambda }}\text{, }\tau
_{0}=a_{1}\left| t_{0}\right| \text{,}
\end{equation}

\begin{equation}
e^{-2\phi }=C^{-1}\exp (-\frac{\tau }{\tau _{0}})\text{.}
\end{equation}

The exact expression for the curvature is 
\begin{equation}
R=\frac{2}{\tau _{0}^{2}}[1+\exp (-\frac{\tau }{\tau _{0}})+\exp (-\frac{%
2\tau }{\tau _{0}})]\text{.}
\end{equation}

Thus, we have a remote singularity in the infinite past and the inflationary
stage in an infinite far future.

Next example is the model (\ref{d}) with $C>0$, $H_{0}=0$, $U=\Lambda
e^{-2\phi }$, $0<t<$ $\infty $. Then it follows from (\ref{1100}) that at $%
t\rightarrow 0$, $\tau \rightarrow -\infty $, 
\begin{equation}
a\thicksim (-\tau )^{s}\text{, }s=1+\kappa Cd\text{, }R\backsim \tau ^{-2}%
\text{, }\phi \backsim \ln \left| \tau \right| \text{,}
\end{equation}
at $t\rightarrow \infty $, $\tau \rightarrow \infty $%
\begin{equation}
a\thicksim \tau ^{-1}\exp (-\tau ^{2})\rightarrow 0\text{, }R\thicksim \tau
^{2}\text{, }\phi \backsim -\ln \tau \text{.}
\end{equation}

Thus, the universe starts from the flat spacetime and infinite scale factor
and exhibits power contraction, at far future we have the remote singularity.

\subsubsection{Type $[C,\Lambda ](\delta ,0)$:}

\begin{eqnarray}
&&e^{CH}=e^{CH_{0}}\left| 1+DCe^{-\delta t}\right| \text{, }g=e^{-\eta }%
\frac{[e^{C(H-H_{0})}\nu -1]}{DC}e^{-2CH}\text{,}  \label{1110} \\
&&R=\frac{U}{(1-2\kappa C)}\{2C+\frac{\omega }{H^{\prime }}+\frac{1}{%
CH^{\prime }}\left( \frac{\omega }{H^{\prime }}\right) ^{\prime }[\nu
e^{C(H-H_{0})}-1]\}\text{,}  \nonumber \\
&&\frac{\Lambda }{(1-2\kappa C)}=-e^{2CH_{0}}D\delta ^{2}\text{, }  \nonumber
\\
&&T_{0}^{0(PL)}=\frac{UC}{4\pi }\frac{\kappa }{1-2\kappa C}(\frac{\omega }{%
CH^{\prime }}+2)\{2+\frac{\omega }{CH^{\prime }}[1-\nu e^{C(H-H_{0})}]\}%
\text{,}  \nonumber
\end{eqnarray}
$\nu $=sign$(1+DCe^{-\delta t})$.

Example

Let us take the same model as in (\ref{a0}) and choose $D=1$, $C>0$, $\delta
<0$, $\frac{\exp (-CH_{0})-1}{C}\equiv \frac{\exp (Ce^{\left| \delta \right|
t_{0}}-1)}{C}\equiv \exp (\left| \delta \right| t_{1})$. Then $%
t_{1}<t<\infty .$ At $t\rightarrow t_{1}$ the solution has the same
asymptotic behavior as for $C=0$ that is deflation $a\backsim \exp (-\tau )$%
. However, behavior at $t\rightarrow \infty $ changes drastically as
compared to $[0,\Lambda ](\delta ,\alpha )$. Indeed, now we have $H\backsim
C^{-1}\left| \delta \right| t$ instead of $H\backsim e^{\left| \delta
\right| t}$. We have, instead of the flat spacetime (inherent to $C=0$ case)
the singularity at the finite value of $\tau =\tau _{0}$, where 
\begin{equation}
g\backsim t^{-1}\exp (-\left| \delta \right| t)\text{, }a\backsim \tau
_{0}-\tau \rightarrow 0\text{, }R\backsim t^{-1}\exp (\left| \delta \right|
t)\text{ }\backsim (\tau _{0}-\tau )^{-2}\ln ^{-2}(\tau _{0}-\tau )\text{.}
\end{equation}

One more example.

$\delta <0$, $D=1$, $C=-\left| C\right| $, $H_{0}=0$, $H$ is the same as in (%
\ref{a0}). Consider the cosmological solution 
\begin{equation}
\exp (-\left| C\right| H)=\exp (\left| \delta \right| t)-1\text{,}
\end{equation}
\begin{equation}
g=\frac{\exp (\left| \delta \right| t)}{U[\exp (\left| \delta \right|
t)-1]^{2}}\text{,}
\end{equation}
defined in the region $0<t<t_{0}$, where $\exp (\left| \delta \right|
t_{0})=2$. Then at $t\rightarrow t_{0}$ $e^{-\phi }\backsim t_{0}-t$, $%
a\backsim \left( t_{0}-t\right) ^{-1}\backsim \exp (\tau )\rightarrow \infty 
$, so we have inflation. At $t\rightarrow 0$ $e^{-2\phi }\backsim -\ln t$, 
\begin{equation}
a\backsim \frac{1}{t\sqrt{(-\ln t)}}\backsim (-\tau )^{-1}\exp (\tau ^{2})%
\text{, }\tau \rightarrow -\infty \text{, }R\backsim \tau ^{2}\text{.}
\end{equation}
Thus, we have the remote singularity in an infinite past.

The type of solutions under discussion possesses one more interesting
pecularity. Usually, in the standard inflationary cosmology, the scalar
field is assumed to be approaching constant, the corresponding effective
potential playing the role of a cosmological constant. Now we will see that
the family of solutions described above contains qualitatively different
possibility: $\left( \nabla \phi \right) ^{2}\neq 0$, but de Sitter
space-time (dS) is an {\it exact} solution of field equations. If the
dilaton depends only on time, so does the metric and we get an exponentially
growing scale factor.

Indeed, let $\omega =\eta =0$, $U=\Lambda =const$. For example, according to
(\ref{exact}), $F=e^{\phi }$, $V=Ce^{2\phi }$ . Then it follows from (\ref
{1110}) that for the solution $[C,\Lambda ](\delta ,0)$ the Riemann
curvature 
\begin{equation}
R=\frac{2\Lambda C}{1-2\kappa C}=-2e^{CH_{0}}DC\delta ^{2}  \label{rc}
\end{equation}
is a constant. If $\Lambda C>0$, we obtain $R>0$, so we have 2D dS metric.
Let, for definiteness, $\delta =-\left| \delta \right| <0$. Then in a remote
past [$1+DC\exp (-\delta t)]>0$. Making a proper shift in time and choosing $%
H_{0}=0$, we may achieve $DC=-1$, $R=2\delta ^{2}$. Then for $t<0$ we have
from (\ref{1110}) 
\begin{equation}
e^{CH}=(1-e^{\left| \delta \right| t})\text{, }g=\frac{e^{\left| \delta
\right| t}}{(1-e^{\left| \delta \right| t})^{2}}\text{, }
\end{equation}
Integrating the expression for $\tau =\int dta$, we see that $t=0$
corresponds to $\tau \rightarrow \infty $. We obtain the exact expressions
in dimensionless variables 
\begin{equation}
\left| \delta \right| \tau =\ln \left[ \frac{1+\exp \frac{\left| \delta
\right| t}{2}}{1-\exp (\frac{\left| \delta \right| t}{2})}\right] \text{, }%
a=sh\hat{\tau}\text{, }\hat{\tau}=\left| \delta \right| \tau \text{, }0<\tau
<\infty \text{,}  \label{i1}
\end{equation}
the behavior of the dilaton is governed by the equation 
\begin{equation}
\exp (CH)=ch^{-2}(\frac{\hat{\tau}}{2})\text{.}  \label{i2}
\end{equation}

The 2D dS metric can be viewed as the metric on the hyperboloid 
\begin{equation}
u^{2}-v^{2}+w^{2}=1\text{,}  \label{hyp}
\end{equation}
embedded in the three-dimensional space 
\begin{equation}
ds^{2}=du^{2}-dv^{2}+dw^{2}\text{.}  \label{dhyp}
\end{equation}

There are three typical possibilities:

\begin{equation}
u=\sinh x\sinh \tau \text{, }v=\cosh x\sinh \tau \text{, }w=\cosh \tau \text{%
,}  \label{u1}
\end{equation}
\begin{equation}
ds^{2}=-d\tau ^{2}+\sinh ^{2}\tau dx^{2}\text{,}  \label{ds1}
\end{equation}
\begin{equation}
u=\sin z\cosh t\text{, }v=\sin z\sinh t\text{, }w=\cos z\text{,}  \label{u2}
\end{equation}
\begin{equation}
ds^{2}=-dt^{2}\sin ^{2}z+dz^{2}\text{,}  \label{ds2}
\end{equation}
\begin{equation}
u=\frac{\rho ^{2}-t_{1}^{2}-1}{2t_{1}}\text{, }v=\frac{\rho ^{2}-t_{1}^{2}+1%
}{2t_{1}}\text{, }w=\frac{\rho }{t_{1}}\text{, }t_{1}=\pm e^{-\tau _{1}}%
\text{,}  \label{u3}
\end{equation}
\begin{equation}
ds^{2}=-d\tau _{1}^{2}+e^{2\tau _{1}}d\rho ^{2}\text{.}  \label{ds3}
\end{equation}

Usually, it is the metric (\ref{ds3}) which is considered in the theory of
inflation. In the region $\left| w\right| <1$, $\left| u\right| >\left|
v\right| $ the metric (\ref{ds3}) can be reduced to the static form (\ref
{ds2}). This is impossible for the metric (\ref{ds1}), for which $\left|
u\right| <\left| v\right| $, $\left| w\right| >1$. The geometry, described
by (\ref{i1}) is geodesically incomplete since at $\tau \rightarrow 0$ we
have a horizon separating a black hole and cosmological (non-static)
regions. Once the point $\tau =0$ is achieved from positive values, a
systems enters the static region.

Eqs. (\ref{i1}), (\ref{i2}) describe inflationary regime at $\tau
\rightarrow \infty $ provided $CH(\phi )\rightarrow -\infty $. If $\omega
\neq 0$, but $\omega \rightarrow 0$ asymptotically, the inflationary regime
can be considered as an approximation.

\subsubsection{Type [$C$, $\Lambda $]($0$, $\alpha $)}

It can be obtained directly from types I or III by putting $\delta =0$.

\subsubsection{Type [$C$, $0$]($\delta $, $\alpha $):}

\begin{eqnarray}
&&CH=\beta _{\pm }t\text{, }g=\exp (-\eta \mp 2\varepsilon CH\beta _{\pm
}^{-1})\text{, }R=-\frac{\beta _{\pm }^{2}}{C^{2}H^{\prime }}\left( \frac{%
\omega }{H^{\prime }}\right) ^{\prime }e^{\eta \pm 2\varepsilon CH\beta
_{\pm }^{-1}}\text{,}  \label{1011} \\
&&T_{0}^{0(PL)}=\frac{1}{4\pi }\exp (\eta \pm 2\varepsilon \frac{CH}{\beta
_{\pm }})\{\kappa \beta _{\pm }(2+\frac{\omega }{CH^{^{\prime }}})[2(\beta
_{\pm }+\delta )+\frac{\beta _{\pm }\omega }{CH^{^{\prime }}})]+2\alpha
(1-2\kappa C)\}\text{,}  \nonumber
\end{eqnarray}
$\beta _{\pm }$ $=-\frac{\delta }{2}\pm \varepsilon $. The solution [$C$, $0$%
]($0$, $\alpha $) does not bring any qualitative new features and can be
obtained directly from (\ref{1011}) by putting $\delta =0$.

Equivalence between $[0,0](\delta _{1},\alpha _{1})$ and $[C,0](\delta
,\alpha )$, $\delta _{1}=2\varepsilon $, $C=\pm \frac{\alpha _{1}\beta _{\pm
}}{\delta _{1}}$.

The dependence $g(\phi )$ coincides with that for $[0,0](0,0)$ provided $%
\delta ^{2}/\alpha =2\varepsilon C/\beta $. The dependence $\tau (\phi )$
coincides with that for $[0,0](\delta ,\alpha )$, provided $\delta
^{2}/\alpha =2\varepsilon C/\beta $.

\subsubsection{Type [$C,0$]($\delta _{0}$,$\alpha $), where $\delta
_{0}^{2}\equiv -4\alpha C$:}

\begin{equation}
H=H_{0}-\frac{\delta _{0}t}{2C}\text{, }g=e^{-\eta }\text{,}
\end{equation}
\begin{equation}
R=\frac{\alpha }{CH^{\prime }}\left( \frac{\omega }{H^{\prime }}\right)
^{\prime }e^{\eta }\text{,}
\end{equation}

\begin{equation}
T_{0}^{0(PL)}=\frac{\alpha }{4\pi }(2-\frac{\kappa \omega ^{2}}{CH^{\prime 2}%
})e^{\eta }\text{.}
\end{equation}

This dependence of the metric and dilaton on time coincides with $[0,0](0,0)$%
, provided $A=-\delta _{0}/2C$.

The solution [$C$, $0$]($\delta $, $0$) can be obtained from (\ref{1011})
directly by putting $\alpha =0$.

\section{Generic types of cosmological solutions}

Let us now discuss briefly the general case with nonzero parameters $%
C,\Lambda $,$\alpha ,\delta $. It is instructive to observe that asymptotic
behavior of the solutions described in Sec. III, in the most part of cases
is qualitatively similar to that found in the particular cases in the
preceding section. This can be seen from general formulas as follows.

Case I$_{a}$.

At $t\rightarrow 0$, 
\begin{equation}
CH\backsim \ln t\text{, }g\backsim \frac{e^{-\eta }}{t^{2}}\text{,}
\end{equation}
at $t\rightarrow \infty $ we have 
\begin{equation}
CH\backsim t,g\backsim e^{-\eta -2\varepsilon t}\text{,}
\end{equation}
It follows from (\ref{0011}) and (\ref{1100}) that this type of solutions
interpolates between the solution $[C,\Lambda ](0,0)$ and the solution with
the zero cosmological constant $[0,0](\delta _{eff},\alpha )$ (where now $%
\delta _{eff}=2\varepsilon $).

Case I$_{b}$.

At $t\rightarrow \pm \infty $, $H\backsim A_{\pm }t$, where the constants $%
A_{+}=C^{-1}(\varepsilon -\delta /2)$, $A_{-}=-C^{-1}(\varepsilon +\delta
/2) $ and, correspondingly, $g\backsim \exp (-\eta +2\varepsilon t)$ and $%
g\backsim \exp (-\eta -2\varepsilon t)$. Thus, at beginning we have
effectively the type $[0,0](\delta _{eff}$, $\alpha _{eff})$ with $\delta
_{eff}=-2\varepsilon $, $\alpha _{eff}=-2\varepsilon A_{-}$ and, at the end,
the same type with $\delta _{eff}=2\varepsilon $, $\alpha
_{eff}=2\varepsilon A_{+}$.

Consider the example with the model (\ref{d}), $\omega =-2$. Let $C<0$, $%
\alpha <0$, so $\varepsilon >\delta /2$. Then $A_{+}<0$, $A_{-}>0$. Both at $%
t\rightarrow -\infty $ and $t\rightarrow \infty $ the dilaton value $\phi
\rightarrow \infty $, $a\backsim \left| \tau \right| $, $R\rightarrow 0$, so
the solution is everywhere regular.

Case II$_{a}$. Then 
\begin{equation}
CH=-\frac{\delta }{2}t+\ln \left| t\right| \text{, }g=\frac{e^{-\eta }}{t^{2}%
}\text{.}
\end{equation}
At $t\rightarrow 0$, again, the behavior is close to that for $[C,\Lambda
](0,0)$. For $t\rightarrow \infty $ the dependence $\phi (t)$ is close to
that for $[0,0](0,0)$ but the dependence $t(\tau )$ may essentially differ
because of the factor $t^{-2}$ in $g(t)$.

Example.

$H=e^{2\phi }-Ae^{-\phi }$, $A>0$, $\delta =-\left| \delta \right| <0$, $%
\omega =-2$.

At $t\rightarrow 0$ 
\begin{equation}
a\backsim \frac{1}{t\left| \ln t\right| }\backsim \exp (\tau +e^{-\tau })%
\text{, }\tau \rightarrow -\infty \text{, }e^{-\phi }\backsim \left| \ln
t\right| \backsim e^{-\tau },\phi \backsim \tau \text{, }R\backsim e^{-2\tau
}\text{.}
\end{equation}

At $t\rightarrow \infty $%
\begin{equation}
a\backsim \frac{1}{\sqrt{t}}\backsim \frac{1}{\tau }\text{, }R\backsim \tau
^{-2}\rightarrow 0\text{, }e^{2\phi }\backsim t\backsim \tau ^{2}.
\end{equation}

Thus, we have the transition from a remote singularity in the past to the
superdeflation in future. If $\delta >0$, in the interval $-\infty <t<0$ we
obtain, in a similar way, the transition from the superinflation in the past
to the remote singularity in the future.

Case II$_{b}$ is equivalent to $[C,0](\delta _{0},\alpha )$.

Case III$.$

Let $\pi n\equiv \varkappa t_{n}\leq \varkappa t\leq \pi (n+1)=$, $\varkappa
t_{n+1}$, $n$ is an integer. As $t\rightarrow t_{k}$, where $k=n$ or $k=n+1$%
, the dependence 
\begin{equation}
CH=const+\ln \left| (t-t_{k})\right| \text{, }a\backsim \frac{\exp (-\eta /2)%
}{\left| t-t_{k}\right| }
\end{equation}
is qualitatively close to that for $[C,\Lambda ](0,0)$.

Thus, we see that, indeed, there is only several typical asymptotic formulas
for time behavior, the behavior of the general solution being described via
its particular cases. However, this standard asymptotic formula are very
sensitive to the choice of a particular model, as we saw it in concrete
examples.

\section{Regular solutions with infinitely strong backreaction}

It is common belief that quantum effects can destroy classical singularity,
thus leading to everywhere regular spacetime. This was confirmed explicitly
for the models considered in \cite{rey}, \cite{bcos}, where it was shown
that it is the finite quantum parameter $\kappa $ that removes the
singularity, inherent to the classical limit $\kappa =0$. In the present
paper, where also examples with everywhere bounded curvature were found, it
is essential that (at least in some cases) this behavior is due to the
nonzero coefficient $d$ or $\kappa $ (they appear as a product $d\kappa $ in
the model (\ref{d})). Thus, quantum backreaction can give rise to regularity
of the metric.

Meanwhile, in 2D dilaton gravity there exist cases when this backreaction
becomes so strong that the contribution from $T_{\mu }^{\nu (PL)}$ diverges
by itself. One could expect these divergencies to destroy regularity of the
spacetime: in this sense not only the absence of backreaction but also too
strong backreaction would seem incompatible with the regularity of the
geometry. Nevertheless, sometimes the geometry can remain finite even in
spite of divergencies in $T_{\mu }^{\nu (PL)}$. This was shown for black
hole solutions \cite{away}. For cosmological ones these results could not be
applied directly since the cosmological counterpart of black holes
considered there would have a finite $\tau $ (analogue of the proper length)
and, thus, would be geodesically incomplete. However, now we will see, using
the materials of previous sections, that combination of divergent quantum
stresses with the regular geometry is possible for cosmological spacetimes
as well.

Let us reconsider the solution $[0,0](\delta ,\alpha )$ for the model
discussed after eq. (\ref{0011}), with $p<1$. For $t\rightarrow \infty $, $%
\phi \rightarrow \infty $, $\tau \rightarrow \infty $, $H^{\prime
}\rightarrow -\kappa d=const$, $R\backsim \exp (-q\phi )$, $%
T_{0}^{0(PL)}\backsim $ $\exp [(2-q)\phi ]$. If $q=2n+2-2p\geq 0$ but $q<2$,
the curvature remains bounded but the energy density of quantum fields
diverges. This is indeed possible and is compatible with the condition of
the geodesic completeness, provided $n<p<1<n+1$.

Let us now turn to the type $[0,\Lambda ](0,\alpha )$. It follows from eq. (%
\ref{0101}) and next formulas that in the limits $\tau \rightarrow \pm
\infty $ , when $\phi \rightarrow -\infty $, $T_{0}^{0(PL)}\backsim
U\backsim \exp (-2\phi )$ becomes infinite. Meanwhile, the metric approaches
de Sitter one with the constant curvature.

The fact that in an infinite past (or infinite future) quantum effects could
be enormously large, are not in disagreement in intuitive expectations about
possible contribution from quantum effects near the classical singularity.
We see that these effects cannot (at least, for some kinds of models)
destroy the regular character of the spacetime.

\section{Summary}

We enumerated all possible types of exact solutions which we have manage to
find among 2D dilaton semiclassical gravity theories with backreaction of
quantum conformal fields taken into account. We analyzed string-inspired
models of dilaton cosmology, in which all coefficients are simple
combinations of exponential and linear functions of the dilaton $\phi $. We
showed that practically any subset of solutions may describe a
singularity-free universe, which starts in an infinite past and expands or
contracts forever. We found also that in some cases these kinds of solution
can occupy the intermediate place between completely regular and singular
spacetimes, representing remote singularities. This means that the
singularity exists but it lies in an infinite past (or infinite far future)
with respect to any event, so not an observer hits it.

Below we summarize main concrete results of our paper. Not repeating
detailed features of solutions, outlined in the main text, we represent in
the table qualitative nature of corresponding initial ($\tau \rightarrow
-\infty $) and final ($\tau \rightarrow \infty $) states. We included in the
table only cases when spacetimes are regular or contain a remote
singularity. This list is by no means exhaustive and represents only some
simple examples of models from string-inspired cosmology. Specifying the
model, we list, for convenience, $H(\phi )$ but it is worth reminding that
the coefficient $F$ can be easily recovered from $H$ according to (\ref{h})$%
. $

$,$%
\begin{tabular}{|l|l|l|l|}
\hline
Type of solution & Model & Initial state & Final state \\ \hline
$\lbrack 0,0](0,0)$ & $H=e^{-2\phi }-\kappa d\phi $, $\omega =-2n$, $0<n<2$
& superinflation & FRW $a\backsim \tau $ \\ \hline
$\lbrack 0,0](0,0)$ & $H=e^{-2\phi }-\kappa d\phi $, $\omega =-4$ & inlfation
& FRW $a\backsim \tau $ \\ \hline
$\lbrack 0,0](\delta ,\alpha )$ & $H=e^{-2\phi }-\kappa d\phi $, $\omega =-2$%
, $\delta ^{2}<2\mu d$ & FRW $a\backsim -\tau $ & FRW $a\backsim \tau $ \\ 
\hline
$\lbrack 0,\Lambda ](\delta ,0)$ & $H=Ae^{-\phi }+e^{-2\phi }\text{, }A>0$, $%
\omega =-2$, & deflation & flat \\ \hline
$\lbrack 0,\Lambda ](\delta ,\alpha )$ & $H=e^{-2\phi }$, $\omega =-2$ & 
deflation & flat \\ \hline
$\lbrack 0,\Lambda ](0,\alpha )$ & $H=e^{2\phi }-Ae^{-2\phi }$, $\omega =-2$
& inflation & deflation \\ \hline
$\lbrack C,\Lambda ](0,0)$ & $H=e^{-2\phi }-\kappa d\phi $, $\omega =-2$ & 
flat & remote singularity \\ \hline
$\lbrack C,\Lambda ](\delta ,0)$ & $H=e^{-2\phi }+Ae^{-\phi }$, $\omega =-2$
& remote singularity & inflation \\ \hline
II$_{a}$ & $H=e^{2\phi }-Ae^{-\phi }$, $A>0$, $\omega =-2$ & remote
singularity & superdeflation \\ \hline
\end{tabular}

Comparing our results with previous studies of exact solutions in 2D
cosmology, we note that, in our terms, the paper \cite{bcos} was devoted to
the types $[0,0](0,0)$ and $[0,0](\delta ,\alpha )$ of solutions where the
model (\ref{d}) was exploited and it was shown that the $[0,0](0,0)$ is
everywhere regular, while $[0,0](\delta ,\alpha )$ contains the singularity.
We saw, however, that the type $[0,0](\delta ,\alpha )$ also may give
singularity-free solutions, provided the signs of the parameters are chosen
properly. Moreover, completely regular solutions exist not only for these
kinds of solutions but practically for all others as well. In so doing, the
violation of WEC (observed in \cite{bcos} for the corresponding particular
case) is the property of some particular types of solutions (such as $%
[0,0](0,0)$ and $[C,\Lambda ](0,0)$), so it seems not to be a necessary or
sufficient condition of the regularity for the generic case. As far as the
property $R>0$ (accelerating solutions) is concerned, \cite{bcos}, this is
fulfilled for $[0,0](0,0)$ and $[0,0](\delta ,\alpha )$ types. However, in a
generic case there are no reasons to expect this property to hold during all
time of evolution. For example, we obtained for the $[0,\Lambda ](0,\alpha )$
type that universe starts from $a=0$, passes through the maximum value of $a$
and ends up with $a=0$. In the beginning and end of evolution $R\backsim 
\frac{\partial ^{2}a}{\partial \tau ^{2}}>0$, but in the point of the
maximum of the scale factor and some its vicinity this quantity is negative,
so the curvature changes its sign somewhere, remaining bounded. In a similar
way, the property of having zero energy density for the classical part of
the effective stress-energy tensor, found in \cite{kim2}, is rather property
of the particular type $[0,0](0,0)$ of solutions, considered there and does
not have to hold in general for all other regular solutions.

Quantum effects are crucial for the most part of the considered examples
since they are responsible for the appearance of terms like $\kappa d$ in
the expression for $H(\phi )$ (see the table) that affects qualitatively the
behavior of the system at $\phi \rightarrow \infty $ or $\phi \rightarrow
-\infty $ in the initial or final state. In this sense, quantum backreaction
turns out to be a powerful tool of removing singularities inherent to
classical solutions. Moreover, in some cases even divergencies of quantum
stresses in the initial or final state ($\tau \rightarrow -\infty $ or $\tau
\rightarrow \infty $) do not spoil the regularity of the geometry. For some
particular types, such as $[0,\Lambda ](0,\alpha )$, the solution does not
exist at all without account for quantum terms.

It is worth paying attention to the non-standard version of the inflation
scenario which is contained in the type $[C,\Lambda ](0,0)$. Being
geodesically incomplete, it represents the part of the de-Sitter world that
expands exponentially fast asymptotically, but in doing so the dilaton field
also depends on time, while the effective potential $U(\phi )$ (which
usually plays a role of a cosmological term with $\phi =const$) is absent.

We restricted ourselves to simplest models. However, the approach, once the
condition of exact solvability is respected, applies to any kind of models
within this class, so the results admit further extension.





%
%

%
%


\begin{references}
\bibitem{callan}  C. G. Callan, S. Giddings, J. A. Harvey, and A.
Strominger, Phys. Rev. D{\bf \ }45 , R1005 (1992); T.Banks, A. Dabholkar, M.
R. Douglas, and M. O. 'Loughlin, Phys. Rev.{\bf \ }D{\bf \ }45, 3607 (1992).

\bibitem{dv}  D. Grumiller, W. Kummer and D. V. Vassilevich, Phys.Rept. 369
(2002) 327.

\bibitem{od}  S.Nojiri and S. Odintsov, Int. J. Mod. Phys. A 16, 1015 (2001).

\bibitem{bil}  A. Bilal and C. G. Callan, Nucl. Phys. B 394, 73 (1993).

\bibitem{alw}  S. P. de Alwis, Phys. Rev. D 46, 5429 (1992).

\bibitem{rst}  J. G. Russo, L. Susskind, and L. Thorlacius, Phys. Rev. D{\bf %
\ }46 (1992) 3444; Phys. Rev. D{\bf \ }47 (1992) 533.

\bibitem{bose}  S. Bose, L. Parker, and Y. Peleg, Phys. Rev. D 52 (1995)
3512.

\bibitem{rob}  G. Michaud and R. C. Myers, Two-Dimensional Dilaton Black
Holes, gr-qc/9508063.

\bibitem{fub}  A. Fabbri and J. G. Russo, Phys. Rev. D 53, 6995 (1995).

\bibitem{kaz}  Y. Kazama, Y. Satoh, and A. Tsuichiya, Phys. Rev. D 51 (1995)
4265.

\bibitem{exact}  O. B. Zaslavskii, Phys. Rev. D 59 (1999) 084013.

\bibitem{class}  O. B. Zaslavskii, Classification of static and homogeneous
solutions in exactly solvable models of two-dimensional dilaton gravity,
hep-th/0212298. To appear in Int. J. Mod. Phys. A.

\bibitem{maz}  F. D. Mazzitelli and J. G. Russo, Phys. Rev. D 47 (1993) 4490.

\bibitem{rey}  S.-J. Rey, Phys. Rev. Lett. 77 (1996) 1929.

\bibitem{rey2}  S.-J. Rey, Recent progress in string inflationary cosmology,
hep-th/9609115.

\bibitem{b1}  S. Bose, Solving the graceful exit problem in superstring
cosmology, hep-th/9704175

\bibitem{bcos}  S. Bose and S. Kar, Phys. Rev. D 56 (1997) 4444.

\bibitem{kim}  W. T. Kim and M. S. Yoon, Phys. Rev. D 58 (1998) 08014.

\bibitem{pl}  A. M. Polyakov, Phys. Lett. 103 B (1981) 207.

\bibitem{kim2}  W. T. Kim and M. S. Yoon, Phys. Lett. B 423 (1998) 231.

\bibitem{away}  O. B. Zaslavskii, Phys. Rev. D 61 (2000) 64002.
\end{references}
\end{document}